\documentclass[pra, twocolumn, showpacs, preprintnumbers, amsmath, amssymb, superscriptaddress]{revtex4-1}
\usepackage{graphicx}
\usepackage{dcolumn}
\usepackage{bm}
\usepackage[bookmarks, colorlinks, linkcolor=blue, citecolor=blue, urlcolor=blue]{hyperref}

\begin{document}

\title{Momentum distribution of atoms in one-dimensional\\Raman cooling with arbitrary coherent population transfer}

\author{Vladimir S. Ivanov}
\email{ivvl82@gmail.com}
\affiliation{Turku Centre for Quantum Physics, Department of Physics and Astronomy, University of Turku, 20014 Turku, Finland}

\author{Kalle-Antti Suominen}
\email{Kalle-Antti.Suominen@utu.fi}
\affiliation{Turku Centre for Quantum Physics, Department of Physics and Astronomy, University of Turku, 20014 Turku, Finland}

\date{\today}

\begin{abstract}
We derive an exact and analytical form for the cold-atom momentum distribution after a large number of one-dimensional (1D) Raman cooling cycles has been applied. Our result shows that one can select pulse profiles and lengths rather freely in order to obtain efficient cooling. Our approach takes into account optical pumping (resonant excitation followed by spontaneous emission) as an intergral part of the process, and it is noted that it does not affect the final momentum distribution. Thus subrecoil cooling is possible in general, and not just for carefully selected combinations of pulse forms and durations.
\end{abstract}


\maketitle

\section{Introduction}

Cold atoms are a fundamental tool for studying degenerate quantum gases~\cite{Weidemuller2009}, ultracold particle collisions~\cite{Suominen1996b,Weiner1999}, many-body effects~\cite{Bloch2008}, quantum computation~\cite{Cirac1995,Haffner2008} and entanglement~\cite{Roghani2008,Roghani2011,Roghani2012}, which then attracts attention to development of various cooling techniques. Laser cooling in combination with evaporative cooling~\cite{Ketterle1996} has made it possible to obtain Bose-Einstein condensation~\cite{Pethick2008}, and cooling of fermionic atoms via sympathetic cooling~\cite{DeMarco1999}. When quantum degeneracy is not required, laser cooling provides large numbers of cold atoms for a reasonable signal-to-noise ratio in atomic clocks~\cite{Wynands2005,Boyd2007}, as well as for unperturbed atomic transition frequencies in ultraprecise measurements~\cite{Ye2008}.

Raman~\cite{Kasevich1992} and \mbox{VSCPT}~\cite{Aspect1988} cooling have led to temperatures below the one-photon recoil limit in one-dimension (1D), but further experiments in 2D~\cite{Davidson1994,Lawall1994} and 3D~\cite{Davidson1994,Lawall1995} have certainly demonstrated the necessity for optimization. Also, proper design of state-insensitive traps is needed for quantum state engineering and precision metrology~\cite{Ye2008,Phoonthong2010,Derevianko2011}. In order to increase Raman-cooling efficiency, a tripod-type configuration of atomic levels was suggested for 2D cooling in Ref.~\cite{Ivanov2011,Ivanov2012b}, as well as variants of coherent population transfer, based on Blackman~\cite{Kasevich1992,Davidson1994,Reichel1994}, square~\cite{Reichel1995,Boyer2004} or \mbox{STIRAP}~\cite{Ivanov2012a,Ivanov2012b} pulses. A comparative analysis of subrecoil cooling is problematic, because it should predict the result of a large number of cooling cycles. On the other hand, even a qualitative analysis~\cite{Reichel1995} given in the L\'evy-flight approach~\cite{Bardou1994} improved an earlier experiment~\cite{Reichel1994} with a better fraction of atoms in the cold peak and a considerably simpler pulse sequence. In turn, our work reported in this paper provides a simple analytical solution for 1D Raman cooling.

Although single cycles in Raman cooling are usually well-defined, there is a large number of them, so that the cooling result becomes almost unpredictable. The coherent population transfer via Raman cycles, implemented already in many ways~\cite{Kasevich1992,Reichel1995,Ivanov2012a}, can be obtained in numerous ways and it is not efficient to test them one by one. Thus the possibility to evaluate the cooling efficiency quickly and in an simple manner is required. In addition, the efficiency of the cooling is affected by spontaneous emission which involves a momentum broadening of the cold atomic ensemble. This aspect must be accounted for when looking for efficient pulse properties.

This paper is organized as follows. In Sec.~\ref{sec:first-step} we define the first step of a Raman cooling cycle with most general the transfer probability. The second step is described in Sec.~\ref{sec:second-step} with an arbitrary angular distribution of spontaneous emission. We show that an analytical solution of the steady-state momentum distribution exists in general, first by deriving in Sec.~\ref{sec:huge-number} an expression for it after a large number of cooling cycles have been applied. The solution itself is derived in the case of most effective transfer profiles in Sec.~\ref{sec:eff-excitation}, and we conclude our presentation with a short discussion on the approach and its implications in Sec.~\ref{sec:conclusion}.

\section{The first step of cooling cycle}
\label{sec:first-step}

Figure~\ref{fig:lambda} shows a three-level $\Lambda$-type atom that travels along the $Oz$ axis, being prepared in the ground state $|1\rangle$. The momentum distribution of atoms is given by the probability amplitude $a(p)$ of the atomic momentum $p = p_z$. The $|1\rangle$ state is coupled by a pump laser beam to the upper state $|2\rangle$, which in turn is coupled to the adjacent ground state $|3\rangle$ by a Stokes laser beam. The first step of a cooling cycle consists of a momentum-selective transfer accomplished through the two-photon transition $|1\rangle \leftrightarrow |3\rangle$. Defining the probability $f(p)$ of the transfer, the number $\Delta N$ of transferred atoms is given by
\begin{align}
  \label{eq:step1-dN}
  \Delta N = f(p) |a(p)|^2,
\end{align}
so that the number of atoms left in the $|1\rangle$ state are
\begin{align}
  \label{eq:step1-a1}
  |a_1(p)|^2 = (1 - f(p)) |a(p)|^2.
\end{align}
Note that spontaneous decay from the upper level $|2\rangle$ is neglected here. Numerous studies of different pulse envelopes such as Blackman~\cite{Kasevich1992,Davidson1994,Reichel1994}, square~\cite{Reichel1995,Boyer2004} and, recently, \mbox{STIRAP}~\cite{Ivanov2012a,Ivanov2012b} pulses, have shown that the undesirable contribution of spontaneous decay during typical pulse durations is suppressed by the large frequency offsets of both laser beams from one-photon resonances. Thereby only the two-photon excitation is included and thus the width of resonant-momentum group is not limited by the upper-state natural linewidth, making a deep subrecoil cooling feasible. For denser samples one needs to be concerned about collisional effects~\cite{Suominen1996b,Holland1994a,Holland1994b} and photoassociation resonances~\cite{Weiner1999}, but these can be avoided e.g. by detuning the laser fields to the blue side of the transition~\cite{Suominen1995,Burnett1996} as shown in Fig.~\ref{fig:lambda}.

The number of lasers included into each laser beam as well as time-dependent changes in beam properties are not specified here. We are only interested in their contribution to the transfer momentum profile which is described by the function $f(p)$. Independently whether a laser beam is provided by several lasers or not, let us assign it an overall wave vector: $\mathbf{k}_1$ for the laser beam near-resonant with the $|1\rangle \leftrightarrow |2\rangle$ transition, and $\mathbf{k}_2$ for the one near-resonant with the $|2\rangle \leftrightarrow |3\rangle$ transition. In ganeral, Raman cooling uses one laser beam configuration to cool atoms with $p<0$, and another one to cool atoms with $p>0$. The first configuration is defined by $\mathbf{k}_1 = k_1 \mathbf{e}_z$, $\mathbf{k}_2 = -k_2 \mathbf{e}_z$ (shown in Fig.~\ref{fig:lambda}), and the second one by $\mathbf{k}_1 = -k_1 \mathbf{e}_z$, $\mathbf{k}_2 = k_2 \mathbf{e}_z$. Depending on the configuration, atoms transferred to the $|3\rangle$ state gain a momentum shift of $\pm\hbar (k_1 + k_2)$, so that, in accordance with Eq.~\eqref{eq:step1-dN}, the momentum distribution in state
$|3\rangle$ is given by
\begin{align}
  \label{eq:step1-a3}
  |a_3(p)|^2 = f(p \mp \hbar (k_1 + k_2)) |a(p \mp \hbar (k_1 + k_2))|^2.
\end{align}
Here, signs ``$-$'' and ``$+$'' correspond to the first and second laser beam configuration, respectively.

\section{The second step of cooling cycle}
\label{sec:second-step}

\begin{figure}
  \includegraphics[width=3.6cm]{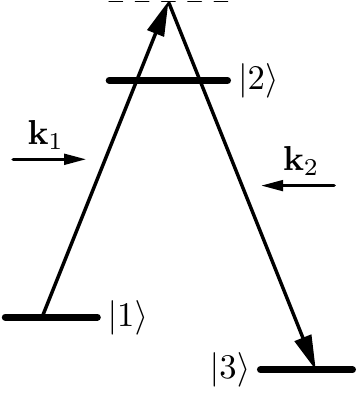}
  \caption{A pump laser beam of wave vector $\mathbf{k}_1$ couples transition $|1\rangle \leftrightarrow |2\rangle$ of a three-level $\Lambda$-type atom, while a Stokes laser beam of $\mathbf{k}_2$ couples the $|2\rangle \leftrightarrow |3\rangle$ transition. A population transfer from the $|1\rangle$ to $|3\rangle$ ground state occurs through a two-photon transition.}
  \label{fig:lambda}
\end{figure}

The second step of a cooling cycle returns atoms into the $|1\rangle$ state by optical pumping. This is done by switching the pump laser in Fig.~\ref{fig:lambda} off and tuning the Stokes laser into resonance with the $|2\rangle \leftrightarrow |3\rangle$ transition. As an atom is excited from $|3\rangle$ to the upper state, its momentum $p'$ is shifted by $\hbar k_2$, becoming $p' \mp \hbar k_2$, with the sign of change depending on the laser beam configuration. Then the atom decays into the $|1\rangle$ state and a spontaneously emitted photon carries away a momentum kick of $\hbar k_1 u$, modifying the atomic momentum as $p = p' \mp \hbar k_2 - \hbar k_1 u$, where $u$ is a random value within range $[-1,1]$.

Consequently, the atomic population in state $|1\rangle$ after this step takes the form
\begin{align}
  |a'(p)|^2 = |a_1(p)|^2 + \int_{-1}^1 N(u) |a_3(p \pm \hbar k_2 + \hbar k_1 u)|^2 du,
\end{align}
where $N(u)$ is the angular distribution of spontaneously emitted photons. With the help of Eqs.~\eqref{eq:step1-a1} and~\eqref{eq:step1-a3}, a change in the atomic population for the complete cooling cycle (steps 1 and 2 combined) is given by
\begin{multline}
  |a'(p)|^2 = (1 - f(p)) |a(p)|^2
  \\
  + \int_{-1}^1 N(u) f(p \mp \hbar k_1 + \hbar k_1 u) |a(p \mp \hbar k_1 + \hbar k_1 u)|^2 du.
\end{multline}
Here, the left-hand side includes all the atoms of momentum $p$ after the cooling cycle, whereas the right-hand side consists of two atomic parts: atoms remaining in the original state and those involved in the cooling process. A transferred atom starts from the $|1\rangle$ state at momentum $p_0 = p \mp \hbar k_1 + \hbar k_1 u$, and subsequently experiences the total momentum shift
\begin{align}
  \label{eq:dp-cycle}
  \delta p = p - p_0 = \pm\hbar k_1 (1 \mp u),
\end{align}
where the sign again depends on the laser beam configuration.

\section{Large number of cooling cycles}
\label{sec:huge-number}

To simplify equations, we set the scaling $\hbar k_1 = 1$. So the $n$th cooling cycle is written as
\begin{multline}
  \label{eq:nth-cycle}
  \rho_n(p) - \rho_{n-1}(p) = -f_n(p) \rho_{n-1}(p)
  \\
  + \int_{-1}^1 N(u) f_n(p \mp 1 + u) \rho_{n-1}(p \mp 1 + u) du,
\end{multline}
where $\rho_{n-1}(p)$ and $\rho_n(p)$ are the atomic populations $|a(p)|^2$ before and after the $n$th cycle, respectively; $f_n(p)$ is the transfer probability.

Equation~\eqref{eq:nth-cycle} is of a special role in understanding the Raman cooling properties. As such a relationship is attained, the usual way to solve the issue is to subsequently specify transfer probabilities $f_n(p)$ and angular distribution $N(u)$, and proceed with the investigation in the way of numerical calculations. Instead, we do not continue by specifying $f_n(p)$ and $N(u)$, but consider repeated cooling cycles in general. Conditions when the atomic population $\rho_n(p)$ tends to the steady-state solution $\rho(p)$ as $n \to \infty$ are discussed in Sec.~\ref{sec:eff-excitation}.

To approach the steady-state solution $\rho(p)$, we consider a large number $N$ of cooling cycles, following $N_0$ initial cycles. The sum of Eq.~\eqref{eq:nth-cycle} over $N$ is given by
\begin{multline}
  \label{eq:sum-rho_n}
  \rho_{N_0+N}(p) - \rho_{N_0}(p) = \sum_{n=N_0+1}^{N_0+N} \biggl( -f_n(p) \rho_{n-1}(p)
  \\
  + \int_{-1}^1 N(u) f_n(p \mp 1 + u) \rho_{n-1}(p \mp 1 + u) du \biggr).
\end{multline}
Once $N_0$ is large enough, the atomic population approaches the steady-state solution, i.e., $\rho_n(p) \approx \rho(p)$, so that the left-hand side of Eq.~\eqref{eq:sum-rho_n} goes to zero, and equation~\eqref{eq:sum-rho_n} leads to inequality
\begin{multline}
  \label{eq:sum-rho}
  \sum_{n=N_0+1}^{N_0+N} \biggl( \int_{-1}^1 N(u) f_n(p \mp 1 + u) \rho(p \mp 1 + u) du
  \\
  - f_n(p) \rho(p) \biggr) \approx 0.
\end{multline}
In general, there are $l$ different profiles for the first laser configuration, $F_1(p),\ldots,F_l(p)$, and the same number of profiles for the second laser configuration, $G_1(p),\ldots,G_l(p)$. As a result, the sum~\eqref{eq:sum-rho} is split into $2l$ sums with similar transfer profiles,
\begin{align}
  \label{eq:F_l-sum}
  \sum_n \biggl( \int_{-1}^1 N(u) F_j(p - 1 + u) \rho(p - 1 + u) du - F_j(p) \rho(p) \biggr),
\end{align}
and
\begin{align}
  \label{eq:G_l-sum}
  \sum_n \biggl( \int_{-1}^1 N(u) G_j(p + 1 + u) \rho(p + 1 + u) du - G_j(p) \rho(p) \biggr),
\end{align}
depending on the laser beam configuration. Since $N$ is large enough, the number of cooling cycles in each of $2l$ sums approximately equals $N/(2l)$. After the substitution of Eqs.~\eqref{eq:F_l-sum} and~\eqref{eq:G_l-sum} for $j=1,\ldots,l$ into Eq.~\eqref{eq:sum-rho}, the latter is written as
\begin{multline}
  \label{eq:rho(p)-av}
  \int_{-1}^1 N(u) F(p - 1 + u) \rho(p - 1 + u) du - F(p) \rho(p)
  \\
  + \int_{-1}^1 N(u) G(p + 1 + u) \rho(p + 1 + u) du - G(p) \rho(p) = 0.
\end{multline}
Here, $F(p)$ and $G(p)$ are the average transfer probabilities for the first and the second laser configurations, respectively:
\begin{align}
  \label{eq:Fp-Gp}
  F(p) = \frac{1}{l} \sum_{j=1}^l F_j(p), \quad G(p) = \frac{1}{l} \sum_{j=1}^l G_j(p).
\end{align}
The change from Eq.~\eqref{eq:sum-rho} to Eq.~\eqref{eq:rho(p)-av} is accompanied by change from an approximate equality to a strict one when $N_0$ and $N$ go to infinity.

\section{The most efficient excitation profiles}
\label{sec:eff-excitation}

As follows from Eq.~\eqref{eq:dp-cycle}, a cooling cycle in the first laser beam configuration pushes the momentum of an atom forwards the $Oz$ axis by a positive shift $\delta p$. Thus transfer of atoms from the left-hand side of the momentum distribution causes cooling, and transfer from the right-hand side causes heating. To avoid the undesirable heating process, the transfer profile should preserve atoms with $p \ge 0$ from the transfer, which is true for e.g. a Blackman and \mbox{STIRAP}-pulse envelope. Without focusing on a concrete momentum profile of the pulse, we impose a general condition
\begin{align}
  \label{eq:Fp=0}
  F(p) > 0, \quad \mbox{if $p < 0$}; \quad F(p) = 0, \quad \mbox{if $p \ge 0$},
\end{align}
in relation to the first laser beam configuration. For the second laser beam configuration the corresponding constraint is written as
\begin{align}
  \label{eq:Gp=0}
  G(p) = 0, \quad \mbox{if $p \le 0$}; \quad G(p) > 0, \quad \mbox{if $p > 0$}.
\end{align}

In addition to efficient cooling, constraints~\eqref{eq:Fp=0} and ~\eqref{eq:Gp=0} ensure that all atoms accumulate around the zero momentum. First of all, one can see that atoms within momentum region $(-2\hbar k_1, 2\hbar k_1)$ do not leave the region if conditions in Eqs.~\eqref{eq:Fp=0} and~\eqref{eq:Gp=0} are satisfied. Atoms from the rest of the ensemble are pushed closer and closer to the center of the distribution until they fall into this region (atoms on the left-hand side are pushed by means of the first laser configuration, and atoms on the right-hand side---by means of the second laser configuration). Finally, the whole atomic ensemble will be transferred near the zero momentum, converging into a steady-state profile as the number of cooling cycles goes to infinity. Such a profile is described by the steady-state solution $\rho(p)$, which in turn takes the form
\begin{align}
  \label{eq:rho(p)=0}
  \rho(p) = 0, \quad \mbox{if $p \le -2$ or $p \ge 2$}.
\end{align}
Whereas Eq.~\eqref{eq:rho(p)=0} only ensures that cooling exists, the cold-atom distribution is derived in Appendix~\ref{sec:solution}, where Eq.~\eqref{eq:rho(p)-final} gives the corresponding steady-state solution:
\begin{align}
  \label{eq:rho(p)}
  \rho(p) = \begin{cases}
    0, & p \le -2;
    \\
    \displaystyle \frac{A}{F(p)} \int_{-1 - p}^1 N(u) du, & p \in (-2,0);
    \\
    \displaystyle \frac{A}{G(p)} \int_{-1}^{1 - p} N(u) du, & p \in (0,2);
    \\
    0, & p \ge 2.
  \end{cases}
\end{align}

\begin{figure}
  \includegraphics[width=8cm]{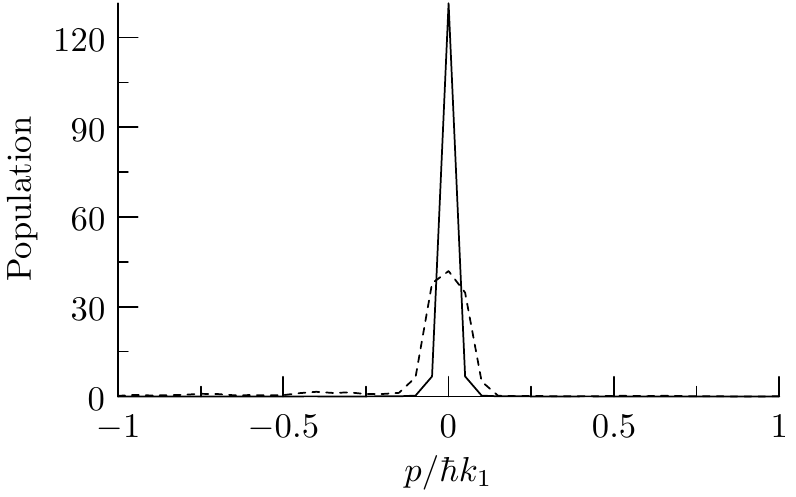}
  \caption{The cold-atom distribution against momentum $p$ evaluated for Raman cooling by \mbox{STIRAP}~\cite{Ivanov2012a}. The analytical result~\eqref{eq:rho(p)} (solid line) shows the attainment of a cooling efficiency superior to that given in the reference (dashed line).}
  \label{fig:cooling}
\end{figure}

As follows from Eq.~\eqref{eq:rho(p)}, the cold-atom distribution can be made arbitrary narrow once average transfer profiles~\eqref{eq:Fp-Gp} have a proper form with a sharp dip at zero momentum. A similar conclusion was pointed out in Ref.~\cite{Ivanov2012a} for a population transfer produced by a \mbox{STIRAP} pulse. A comparison between the resulting cold-atom distribution in Ref.~\cite{Ivanov2012a} and that given by Eq.~\eqref{eq:rho(p)} is shown in Fig.~\ref{fig:cooling}. The analytical result~\eqref{eq:rho(p)} allows even more efficient cooling in comparison with Ref.~\cite{Ivanov2012a}.

\section{Conclusion}
\label{sec:conclusion}

We have presented the general result of 1D Raman cooling as a simple analytical expression~\eqref{eq:rho(p)} which combines a large number $N$ of cooling cycles and gives the steady-state solution when $N\to\infty$. So far, Raman cooling has relied on the set of a specific order of profiles introduced by Kasevich and Chu in 1992~\cite{Kasevich1992}; each profile in the set is narrower and closer to $v=0$ than the previous one in order to transfer most of the momentum distribution. Nevertheless, the resulting cold-atom distribution~\eqref{eq:rho(p)} is defined by averaged excitation profiles~\eqref{eq:Fp-Gp} and is independent of the order of profiles. If needed, not only arbitrary order of excitation profiles can be applied with the same efficiency, but rather different profile envelops such as that produced by a \mbox{STIRAP} pulse~\cite{Ivanov2012a}, combining the excitation of a momentum-distribution wing with a narrow dip at $v=0$. In addition, the steady-state solution is independent of any particular population transfer method, allowing one to observe the influence of spontaneous decay accompanying the second step in each cooling cycle. One can see that spontaneous decay does not impose any constraint on neither the height or width of the cold peak, making deep subrecoil cooling attainable. However, one should note that taking into account spontaneous decay through channel $|2\rangle \leftrightarrow |3\rangle$ gives more realistic situations, but leads to a significantly complicated consideration in the frame of the Wigner function~\cite{Wigner1932}, and we have not done it here.

\section{Acknowledgments}

V. I. would like to thank Yu. V. Rozhdestvensky for helpful discussions. This research was supported by the Finnish Academy of Science and Letters, CIMO, and the Academy of Finland, grant 133682.

\appendix
\section{Steady-state solution}
\label{sec:solution}

The steady-state solution $\rho(p)$ in Eq.~\eqref{eq:rho(p)-av} is now multiplied by either $F(p)$ or $G(p)$ transfer probability. Consequently, one can introduce functions
\begin{align}
  \label{eq:def-y1p-y2p}
  y_1(p) = F(-p) \rho(-p), \quad y_2(p) = G(p) \rho(p),
\end{align}
which both are nonzero over range $p \in [0,2)$ only, as follows from Eqs.~\eqref{eq:Fp=0},~\eqref{eq:Gp=0} and \eqref{eq:rho(p)=0}. Thus, functions $y_{1,2}(p)$ can be nonzero at $p=0$ due to the fact that $\rho(p)$ can go to infinity as $p \to 0$. In terms of variables $y_{1,2}(p)$ Eq.~\eqref{eq:rho(p)-av} is written as
\begin{multline}
  \label{eq:y1p-y2p-rel}
  \int_{-1}^1 N(u) y_1(-p + 1 - u) du - y_1(-p)
  \\
  + \int_{-1}^1 N(u) y_2(p + 1 + u) du - y_2(p) = 0.
\end{multline}

Although Eq.~\eqref{eq:y1p-y2p-rel} depends on two functions, $y_1(p)$ and $y_2(p)$, it can be reduced for $p \in (0,2)$ to an equation of one function
\begin{align}
  \label{eq:def-vp}
  v(p) = y_1(2 - p) + y_2(p).
\end{align}
First, we replace $p$ by $p-2$ in Eq.~\eqref{eq:y1p-y2p-rel}:
\begin{multline}
  \label{eq:y1p-y2p-shifted}
  \int_{-1}^1 N(u) y_1(-p + 3 - u) du - y_1(2 - p)
  \\
  + \int_{-1}^1 N(u) y_2(p - 1 + u) du - y_2(p - 2) = 0,
\end{multline}
and then evaluate the sum of Eqs.~\eqref{eq:y1p-y2p-rel} and~\eqref{eq:y1p-y2p-shifted}:
\begin{multline}
  \label{eq:y1p-y2p-sum}
  \{y_1(2 - p) + y_2(p)\} + \{y_1(-p) + y_2(p - 2)\}
  \\
  = \int_{-1}^1 N(u) \{y_1(-p + 1 - u) + y_2(p + 1 + u)\} du
  \\
  + \int_{-1}^1 N(u) \{y_1(-p + 3 - u) + y_2(p - 1 + u)\} du.
\end{multline}
Excluding $y_1(-p)$ and $y_2(p - 2)$, which both equal zero over $p \in (0,2)$, the rest of Eq.~\eqref{eq:y1p-y2p-sum} is written in terms of function $v(p)$~\eqref{eq:def-vp} as
\begin{multline}
  \label{eq:rel-vp}
  v(p) = \int_{-1}^{1 - p} N(u) v(p + 1 + u) du
  \\
  + \int_{1 - p}^1 N(u) v(p - 1 + u) du,
\end{multline}
where we take into account that $v(p) = 0$ if $p \notin [0, 2]$.

One can see that Eq.~\eqref{eq:rel-vp} depends only on $v(p)$ over range $p \in (0,2)$. That is why if another function $\bar v(p)$ coincides with $v(p)$ for all $p \in (0,2)$, $\bar v(p)$ satisfies Eq.~\eqref{eq:rel-vp} as well. At the same time, $\bar v(p)$ can take arbitrary values if $p \notin (0,2)$. We assume that $\bar v(p)$ is a periodic function with period $T = 2$, hence Eq.~\eqref{eq:rel-vp} applied for $\bar v(p)$ is written as
\begin{align}
  \label{eq:rel-bar-vp}
  \bar v(p) = \int_{-1}^1 N(u) \bar v(p + 1 + u) du.
\end{align}
Here, we use relationship $\bar v(p + T) = \bar v(p)$, and assume that $\bar v(nT) = y_1(0) + y_2(0)$ for all integer $n$.

The periodic function $\bar v(p)$ is expanded into the Fourier series
\begin{align}
  \label{eq:fourier-vp}
  \bar v(p) = \sum_k e^{ik\pi p} a_k.
\end{align}
In terms of coefficients $a_k$, Eq.~\eqref{eq:rel-bar-vp} takes the form
\begin{align}
  \label{eq:rel-ak}
  x_k a_k = 0, \quad x_k = 1 - (-1)^k \int_{-1}^1 N(u) e^{ik\pi u} du.
\end{align}
It is straightforward to verify that
\begin{align}
  x_k = 0, \quad \mbox{if $k = 0$}; \quad x_k \ne 0, \quad \mbox{if $k \ne 0$};
\end{align}
hence, Eq.~\eqref{eq:rel-ak} is equivalent to
\begin{align}
  a_k = A \delta_{k0},
\end{align}
where $A$ is an arbitrary constant and $\delta_{kn}$ is Kronecker's delta. As a result, $v(p)$ over range $p \in (0, 2)$ is given by
\begin{align}
  \label{eq:vp}
  v(p) = A.
\end{align}

If $p \in (0, 2)$, Eqs.~\eqref{eq:y1p-y2p-rel} and~\eqref{eq:y1p-y2p-shifted} are written as
\begin{align}
  y_2(p) = \int_{-1}^{1 - p} N(u) v(p + 1 + u) du,
  \\
  y_1(2 - p) = \int_{1 - p}^1 N(u) v(p - 1 + u) du.
\end{align}
Hence, functions $y_{1,2}(p)$ are given by
\begin{align}
  y_1(p) = A \int_{p - 1}^1 N(u) du, \quad y_2(p) = A \int_{-1}^{1 - p} N(u) du.
\end{align}
Finally, equations in Eq.~\eqref{eq:def-y1p-y2p} give the steady-state
solution
\begin{align}
  \label{eq:rho(p)-final}
  \rho(p) = \begin{cases}
    0, & p \le -2;
    \\
    \displaystyle \frac{A}{F(p)} \int_{-1 - p}^1 N(u) du, & p \in (-2,0);
    \\
    \displaystyle \frac{A}{G(p)} \int_{-1}^{1 - p} N(u) du, & p \in (0,2);
    \\
    0, & p \ge 2;
  \end{cases}
\end{align}
where $F(p)$, $G(p)$ follow from Eq.~\eqref{eq:Fp-Gp}.

\bibliography{Raman-cooling}
\bibliographystyle{apsrev4-1}

\end{document}